\documentclass[aps,twocolumn,showpacs,amsmath,amssymb,pra,superscriptaddress,floatfix,longbibliography]{revtex4-1}
\usepackage{braket}
\usepackage{amsmath}
\usepackage{amssymb}
\usepackage{mathtools}
\usepackage{graphicx}
\usepackage{dcolumn}
\usepackage{bm}
\usepackage{multirow}
\usepackage{leftidx}
\usepackage{color}
\usepackage{float}
\usepackage{xcolor}
\usepackage{soul}

\usepackage{amsfonts}
\usepackage{eufrak}
\usepackage[german,english]{babel}

\begin{document}
\title{Beyond lowest order mean field theory for light interacting with atom arrays}
\author{F.~Robicheaux}
\email{robichf@purdue.edu}
\affiliation{Department of Physics and Astronomy, Purdue University, West Lafayette,
Indiana 47907, USA}
\affiliation{Purdue Quantum Science and Engineering Institute, Purdue
University, West Lafayette, Indiana 47907, USA}
\author{Deepak A.~Suresh}
\affiliation{Department of Physics and Astronomy, Purdue University, West Lafayette,
Indiana 47907, USA}

\date{\today}

\begin{abstract}
Results from higher order mean field calculations of light interacting
with atom arrays are presented for calculations of one- and two-time
expectation values. The atoms are approximated as two-levels and are
fixed in space. Calculations were performed for mean field approximations
that include the expectation value of one operator (mean field),
two operators (mean field-2), and three operators (mean field-3).
For the one-time expectation values, we examined three
different situations to understand the convergence with increasing order
of mean field and some limitations of higher order mean field approximations.
As a representation of a two-time expectation value, we calculated
the $g^{(2)}(\tau )$ for a line of atoms illuminated by a perpendicular plane wave
at several emission angles and two different intensities. For many cases,
the mean field-2 will be sufficiently accurate to quantitatively predict
the response of the atoms as measured by one-time expectation values.
However, the mean field-3 approximation will often be needed for
two-time expectation values.
\end{abstract}

\maketitle


\section{Introduction}

Light interacting with many closely spaced atoms leads to several
interesting and complex many body effects. Typically, these effects
arise when the spacing between atoms is of order the light's wavelength
or smaller. Superradiance and subradiance\cite{RHD1954,RH11971,GH11982,SFO2006,SCS2008,SC12008,KS12014,CKB2018,CPD2011,LY12012,BPK2012,DDY2015,GAK2016,SR32017,OMG2019,AMA2017,ZM12019,GGV2019,NLO2019,MFO2020,RWR2020}
are classic effects whereby the collective interaction with the light field
substantially slows (subradiance) or speeds up (superradiance) the
rate that light is emitted from the atom cloud. In addition, the
collective interactions between the atoms 
can lead to the shift in the resonance
frequency\cite{CYL2004,MSS2014,CES2020,FHM1973,SCS2010,ILB2005,KSK2012,JRL2014,GFS2020,JR12016}
or can qualitatively change the scattering pattern from the atom cloud\cite{CBL2010,RSB2014,PBJ2014,JR12016,BZB2016,JBS2016,ZCY2016,GFS2020,RWR2020,RS12020,SWL2017}.
In recent years, several groups have studied the possibility for using
this collective interaction for the manipulation of
light\cite{JR12012,BGA2016,SWL2017,GGV2018,GGV2019,RWR2020}. One
particularly promising system is for the atoms to be in a regular array
\cite{CYL2004,JR12012,MSS2014,BGA2016,SR22016,SWL2017,AMA2017,GGV2018,QR12019,ZM12019,AKC2019,HDC2019,GGV2019,NLO2019,WBR2020,WR12020,MFO2020,RWR2020,BLG2020,CES2020}
in order to decrease the dephasing from near-field atom-atom interactions which
leads to stronger collective effects and  qualitatively new phenomena.

Calculations of light interacting with a group of atoms have tended
to fall into two groups. If the number of atoms is small enough,
full density matrix calculations are performed. These calculations
can be thought of as being exact in the sense that the results are
progressively more accurate as, for
example, $\delta t$ gets smaller and the number of included states
gets larger. If there are a large number of
atoms and many possible
excitations, mean field calculations are performed because the state space of
the density matrix quickly becomes too large to solve on even the
largest computers, much less the resources of the average
researcher. Often then,
the time dependent operator equations are approximated by replacing
two atom expectation values by the product of one atom expectation
values\cite{BPK2012,JRL2014,MSS2014,JBS2016,JRL2017,SR32017,WR12020,GFS2020,BLG2020,DWC2020}.
This method, the mean field approximation,
has an inherent accuracy in the sense that
for a given situation the error can not be reduced below a fixed,
non-zero value by, for example, decreasing
$\delta t$. The mean field approximation should work well if
there is not much correlation in the system. If the light intensity
is low or if it is known that there is only one excitation
in the system, then the linear approximation, where the pair operators
are set to zero, can be
used\cite{FHM1973,SFO2006,SCS2008,SC12008,SCS2010,CBL2010,JR12012,PBJ2014,RSB2014,KS12014,SS12015,SR12016,SR22016,JR12016,BGA2016,BZB2016,ZCY2016,SWL2017,CKB2018,GGV2018,GGV2019,NLO2019,CES2020,RS12020}. The strengths and weaknesses of
these methods are fairly obvious. The density matrix calculations
are accurate but require inordinate time and memory as the number of
atoms increases. The mean field or linear approximation lose accuracy
as the correlation between atoms increases but can be applied
to many atoms ($10^3$ atoms are routine,
and more than $10^5$ atoms in one case\cite{RS12020}).

There are approximations with accuracy between that of the mean
field approximation and a density matrix treatment\cite{RJ11997}.
Reference~\cite{KR12015}
described a method they called generalized mean-field which
includes higher order correlations and explicitly derived
the time dependent operator equations that correctly include single and
pair expectation values but approximate triple expectation values;
below, we will call this the mean field-2 (MF2) approximation. (Using
this convention, the usual mean field approximation is MF1.)
The MF2 approximation
should be more accurate than the MF1 approximation but
without the computational effort of a density matrix calculation
and has been used in several
calculations\cite{FY11999,CYL2004,LY12012,KR12015,QR12019,OMG2019,HPR2020,KK12018,KRK2019,SSF2020}.
However, it is substantially slower than the MF1
method. For $N$ atoms, the number of
operations for one time step is proportional to $3^1 N^2$ for the
MF1 approximation and is proportional to $3^2 N^3$ for the
MF2 approximation. The extra computational time and extra
programming complexity explains why the MF2
approximation is rarely used in practice.

Even higher
order mean field approximations result from keeping larger products
in the time dependent operator equations. For example, the
MF3 equations correctly include single, pair, and triple
expectation values but approximates quadruple expectation values.
The number of operations for the MF3 approximation
is proportional to $3^3 N^4$ which can be onerous but is much faster than
evaluating the full density matrix.

Besides expectation values of operators, two-time expectation values
are often of interest\cite{FY11999,RL12000,LY12012,JSO2016,CES2020,WBR2020,MFO2020}. For example, the $g^{(2)}(\tau )$ function,
Eq.~(\ref{Eqg2}), is the
normalized intensity-intensity correlation of emitted light.
In principle, this can be calculated
using  higher order mean field approximations. Realizing
this, we adapted the MF2 and MF3 approximation to the calculation
of two-time expectation values. However, unlike the calculation
of expectation values, the MF2 approximation often
did not give accurate results for two time
expectation values. We found a complete
breakdown of the approximation even for very low light
intensities where we expected the small number of excitations
would lead to weak, if any, correlation. In one case, we decreased
the intensity until there was less than $10^{-5}$ total excitations
and the MF2 approximation still failed. For
$g^{(2)}(\tau )$, there can be a strong dependence on 3 or
more atom correlations even for weak excitations.

Spurred by this failure of the MF2 method, we followed the
spirit of the
derivation in Ref.~\cite{KR12015} to include expectation values of
triple atom operators and tested the resulting approximation in a
variety of cases from expectation values in a nearly uncorrelated
system to two-time expectation values in a correlated system. The tests
were strict in the sense that we increased the order of the approximation
from MF1 to MF2 to MF3 until convergence was obtained or we directly
compared
with the results from full density matrix calculations. We found that
the MF3 method often gave accurate results even when the
MF2 method completely fails (more than 100\% error).

The paper is organized as follows: Sec.~\ref{SecBT} presents the
basic equations for approximating expectation values and two-time
expectation values (e.g.~$g^{(2)}(\tau )$), Sec.~\ref{SecR} 
contains results for three situations involving expectation values
and one situation for two-time expectation values,
Sec.~\ref{SecS} summarizes the results, and the appendices
Sec.~\ref{SecAppRed} describes the method for reducing the order
of operators and Sec.~\ref{SecTtIC} describes the method for obtaining
the initial conditions for two-time expectation values.

\section{Basic Theory}\label{SecBT}

We are using an excitation scheme where the atomic structure is approximated
as a two level system. The atoms will be considered as fixed in space which
means we are ignoring the atom recoil.

\subsection{Master equation formalism}

All of the equations will use a
simplified notation to reduce the size of the resulting equations.
For the $n$-th atom, the ground and excited states are $|g_n\rangle$ and
$|e_n\rangle$. The operators used below follow the definition
\begin{equation}
\hat{e}_n\equiv  |e_n\rangle\langle e_n|\qquad
\hat{\sigma}^-_n\equiv |g_n\rangle\langle e_n|\qquad
\hat{\sigma}^+_n\equiv|e_n\rangle\langle g_n|
\end{equation}

The equation for the $N$-atom density matrix\cite{RL12000} can be written in
the form
\begin{eqnarray}\label{EqDenMat}
\frac{d\hat{\rho}}{dt}&=&\sum_n\left[\frac{1}{i\hbar}[H_n,\hat{\rho}]
+{\cal L}_n(\hat{\rho })\right]\nonumber\\
&+&\sum_n\sum_m{\vphantom{\sum}}'\left[\frac{1}{i\hbar}[H_{nm},\hat{\rho}]
+{\cal L}_{nm}(\hat{\rho})
\right]
\end{eqnarray}
where a prime on the sum means $m\neq n$,
the $H_n$ is the one atom Hamiltonian that arises from an external
laser interacting with each atom, ${\cal L}_n$ is from one atom decays
of Lindblad type, the $H_{nm}$ is the two atom Hamiltonian from the
dipole-dipole interactions, and ${\cal L}_{nm}$ are the two atom decays
from the dipole-dipole interactions. For the two level cases considered
here, these operators are
\begin{eqnarray}
H_n&=&\hbar \left(\frac{\Omega^+_n}{2}\hat{\sigma}_n^++\frac{\Omega^-_n}{2}
\hat{\sigma}^-_n-\Delta_n\hat{e}_n\right)\\
{\cal L}_n(\hat{\rho })&=&
\frac{\Gamma}{2}(2\hat{\sigma}_n^-\hat{\rho}\hat{\sigma}_n^+-\hat{e}_n
\hat{\rho}-\hat{\rho}\hat{e}_n)\\
H_{nm}&=&\hbar\Omega_{nm}\hat{\sigma}^+_n\hat{\sigma}_{m}^-\\
{\cal L}_{nm}&=&\frac{\Gamma_{nm}}{2}(2\hat{\sigma}_n^-\hat{\rho}
\hat{\sigma}_{m}^+-\hat{\sigma}_{m}^+\hat{\sigma}_n^-
\hat{\rho}-\hat{\rho}\hat{\sigma}_{m}^+\hat{\sigma}_n^-)
\end{eqnarray}
where $\Delta_n$ is the detuning of the transition for atom-$n$, the
$\Omega^+_n =(\Omega_n^-)^*$ is the complex Rabi frequency at atom-$n$
(for plane wave light, $\Omega_n^+=\Omega \exp [i\vec{k}\cdot\vec{R}_n]$
where $\vec{k}$ is the wave number and $\vec{R}_n$ is the atom position).
The two atom parameters are defined for $m\neq n$ as
\begin{eqnarray}
\Gamma_{nm}&=&g(\vec{R}_{nm})+g^*(\vec{R}_{nm})=
2\Re [g(\vec{R}_{nm})]\label{Eqgdef1}\\
\Omega_{nm}&=&\frac{g(\vec{R}_{nm})-g^*(\vec{R}_{nm})}{2i}=
\Im [g(\vec{R}_{nm})]\label{Eqgdef2}\\
g(\vec{R}) &=&\frac{\Gamma}{2}\left[h_0^{(1)}(s)+
\frac{3\hat{R}\cdot\hat{d}^*\hat{R}\cdot\hat{d}-1}{2}
h_2^{(1)}(s)\right]\label{Eqgdef3}\\
g^\pm_{nm} &\equiv &\pm i\Omega_{nm} + \frac{1}{2}\Gamma_{nm}\label{Eqgdef4}
\end{eqnarray}
with $\hat{d}$ the dipole unit vector,
$s=kR$, $\hat{R}=\vec{R}/R$, and the $h_\ell^{(1)}(s)$ the outgoing spherical
Hankel function of angular momentum $\ell$:
$h_0^{(1)}(s)=e^{is}/[is]$ and $h_2^{(1)}(s) = (-3i/s^3 - 3/s^2 + i/s)e^{is}$. 
The $g(\vec{R})$ is proportional to the propagator that gives the
electric field at $\vec{R}$ given a dipole at the origin\cite{JDJ1999}.
For
a $\Delta M =0$ transition, $\hat{d}=\hat{z}$ and the coefficient
of the $h_2^{(1)}$ Bessel function is $P_2(\cos (\theta ))=
(3\cos^2(\theta )-1)/2$ where $\cos(\theta )=Z/R$. For
a $\Delta M =\pm 1$ transition, the coefficient
of the $h_2^{(1)}$ Bessel function is $-(1/2)P_2(\cos (\theta ))=
(1-3\cos^2(\theta ))/4$.

\subsection{Operator equation formalism}

The developments in this section are similar to those in Ref.~\cite{KR12015}
for including expectation values of two operators with the exception that
Ref.~\cite{KR12015} uses the operators that give all real equations, $\hat{\sigma}_{x,y,z}$,
whereas we use operators that lead to complex equations, $\hat{\sigma}^\pm$.
We have shown that our equations are the same as in Ref.~\cite{KR12015}
except for the typo in their Eq.~(5c) where the $-i$ should be $-1$.

In order to make the notation more compact in this
section, we will define the following
\begin{equation}
\hat{Q}_n^{-1}=\hat{\sigma}^-_n\qquad
\hat{Q}_n^{0}=\hat{e}_n\qquad
\hat{Q}_n^{1}=\hat{\sigma}^+_n\qquad
\end{equation}

\subsubsection{One atom contributions}

This section gives the contribution to the time dependence of the
expectation value of operators due to the terms with $H_j$
and ${\cal L}_j$. The equations of motion for the one atom terms are
\begin{equation}\label{EqOne1}
\frac{d\langle \hat{Q}_n^j\rangle}{dt}=Tr\left[\hat{Q}_n^j\frac{d\hat{\rho}}{dt}
\right] ={\cal S}_n^j +
\sum_{j'=-1}^1W^{jj'}_n\langle \hat{Q}_n^{j'}\rangle
\end{equation}
where the second step results by replacing the derivative of the density
matrix with the first line of the right hand side of Eq.~(\ref{EqDenMat}). Performing
this substitution gives (counting rows and columns in order: $-1, 0, 1$)
\begin{eqnarray}
\underline{W}_n&=&\begin{pmatrix}
i\Delta_n-\frac{\Gamma}{2}& i\Omega^+_n&0\\
i\frac{\Omega^-_n}{2} & -\Gamma & -i\frac{\Omega^+_n}{2}\\
0 & -i\Omega^-_n & -i\Delta_n - \frac{\Gamma}{2}
\end{pmatrix}
\\
\vec{\cal S}_n&=&\begin{pmatrix}
-i\frac{\Omega^+_n}{2} \\ 0 \\ i\frac{\Omega^-_n}{2}
\end{pmatrix}
\end{eqnarray}

Using this notation, the rate of change of the
two atom expectation values due to the one atom terms in
Eq.~(\ref{EqDenMat}) can found from Eq.~(\ref{EqOne1})
and are
\begin{eqnarray}\label{EqOne2}
&\null &\frac{d\langle \hat{Q}_n^j\hat{Q}_{m}^{j'}\rangle}{dt}={\cal S}_n^j
\langle \hat{Q}_{m}^{j'}\rangle + \langle \hat{Q}_n^j\rangle
{\cal S}_{m}^{j'} +\nonumber \\
&\null &\sum_{j''=-1}^1\left[W^{jj''}_n\langle
\hat{Q}_n^{j''}\hat{Q}_{m}^{j'}\rangle
+W^{j'j''}_{m}\langle \hat{Q}_n^j\hat{Q}_{m}^{j''}\rangle\right]
\end{eqnarray}
where $n\neq m$.

The rate of change of the three and higher atom expecation values
can be generalized from these equation.

\subsubsection{Two atom contributions}

The two atom contributions to the rate of change of the one atom expectation
values is
\begin{eqnarray}
\frac{d\langle\hat{Q}^0_n\rangle }{dt}&=&
-\sum_m{\vphantom{\sum}}'(g^+_{nm}\langle \hat{Q}^1_n\hat{Q}^{- 1}_{m}\rangle
+g^-_{nm}\langle \hat{Q}^{- 1}_n\hat{Q}^1_{m}\rangle )\label{EqTwo1} \\
\frac{d\langle\hat{Q}^{\pm 1}_n\rangle }{dt}&=&\sum_m{\vphantom{\sum}}'
g_{nm}^\mp (2\langle \hat{Q}^0_n\hat{Q}^{\pm 1}_{m}\rangle 
-\langle \hat{Q}^{\pm 1}_{m}\rangle )\label{EqTwo2}
\end{eqnarray}
where a prime on the sum means $m\neq n$ and
$g^\pm_{nm}$ are defined by Eq.~(\ref{Eqgdef4}). These equations
can be generalized to the form
\begin{equation}\label{EqTwo1g}
\frac{d\langle\hat{Q}^j_n\rangle }{dt}=
\sum_{j'm}{\vphantom{\sum}}'V^{jj'}_{nm}\langle \hat{Q}^{j'}_m\rangle
+\sum_{j'j''m}{\vphantom{\sum}}'U^{jj'j''}_{nm}
\langle \hat{Q}^{j'}_n\hat{Q}^{j''}_m\rangle
\end{equation}
where a prime on the sum means $m\neq n$ and the only nonzero elements are
\begin{eqnarray}
V^{\pm 1,\pm 1}_{nm}&=&-g^\mp_{nm} \\
U^{0,\pm 1,\mp 1}_{nm}&=&-g^\pm_{nm}\qquad
U^{\pm 1,0,\pm 1}_{nm}= 2g^\mp_{nm}
\end{eqnarray}

In the two atom expectation values, there are terms involving
$H_{nm}$ and ${\cal L}_{nm}$ and there are terms involving
$nl$ and $ml$ subscripts where $l\neq n$ {\it and} $l\neq m$.
The latter terms can be found from application of Eqs.~(\ref{EqTwo1}).
This will lead to 3 operator expectation values in the expressions for
the rate. These equations are
\begin{eqnarray}
\frac{d\langle \hat{Q}_n^j\hat{Q}_{m}^{j'}\rangle}{dt}&=&
\sum_{j''l}{\vphantom{\sum}}'\left[ V^{jj''}_{nl}\langle \hat{Q}_l^{j''}\hat{Q}_{m}^{j'}\rangle
+V^{j'j''}_{ml}\langle\hat{Q}_n^j\hat{Q}_{l}^{j''}\rangle\right]\nonumber \\
&\null &+\sum_{j''j'''l}{\vphantom{\sum}}'U^{jj''j'''}_{nl}\langle \hat{Q}_n^{j''}\hat{Q}_l^{j'''}\hat{Q}_{m}^{j'}\rangle\nonumber \\
&\null &+\sum_{j''j'''l}{\vphantom{\sum}}'U^{j'j''j'''}_{ml}\langle \hat{Q}_n^j\hat{Q}_{m}^{j''}\hat{Q}_l^{j'''}\rangle
\end{eqnarray}
where $m\neq n$ and a prime on the sum means $l\neq m,n$.
The only nonzero contribution from $H_{nm}$ and ${\cal L}_{nm}$
that do not lead to connections with atoms $l\neq m,n$ are
\begin{eqnarray}
\frac{d\langle \hat{Q}^{-1}_n\hat{Q}^{+ 1}_{m}\rangle }{dt}
&=&2\Gamma_{nm}\langle \hat{Q}^0_n\hat{Q}^0_{m}\rangle
-g^+_{nm}\langle \hat{Q}^0_{m}\rangle
-g^-_{nm}\langle \hat{Q}^0_n\rangle\label{EqTwo3} \\
\frac{d\langle \hat{Q}^0_n\hat{Q}^{\pm 1}_{m}\rangle }{dt}
&=& -g^\pm_{nm}\langle \hat{Q}^{\pm 1}_n\hat{Q}^0_{m}\rangle
\end{eqnarray}
where the $\Gamma_{nm}$ are defined in Eq.~(\ref{Eqgdef1}).

As an example that includes both types of terms, the two atom
contributions give
\begin{eqnarray}\label{EqPMder}
\frac{d\langle \hat{\sigma}^+_m\hat{\sigma}^-_n\rangle}{dt}&=&
2\Gamma_{mn}\langle \hat{e}_m\hat{e}_n\rangle -
g^+_{nm}\langle\hat{e}_m\rangle -g^-_{nm}\langle\hat{e}_n\rangle \nonumber\\
&\null&+\sum_l{\vphantom{\sum}}'g^+_{nl}\langle 2\hat{e}_n\hat{\sigma}^-_l\hat{\sigma}^+_m
-\hat{\sigma}^-_l\hat{\sigma}^+_m\rangle\nonumber\\
&\null&+\sum_l{\vphantom{\sum}}'g^-_{ml}\langle 2\hat{e}_m\hat{\sigma}^+_l\hat{\sigma}^-_n
-\hat{\sigma}^+_l\hat{\sigma}^-_n\rangle
\end{eqnarray}
where the first line is from Eq.~(\ref{EqTwo3}) and the terms in the
sum are from Eq.~(\ref{EqTwo2}) applied to the $\hat{\sigma}^-_n$ and
$\hat{\sigma}^+_{m}$ operators.

The rate of change for 3 operator expectation values can be generated
from these equations.

\subsection{Two-time correlation functions}

In many situations, the calculation of two-time correlation functions can
lead to useful information about the system, for example the correlation
in emitted photons. As an example, consider the combination
\begin{equation}\label{Eqg2}
g^{(2)}(\tau )=\lim_{t\to\infty}\frac{
\langle \hat{\sigma}^+(t)\hat{\sigma}^+(t+\tau)\hat{\sigma}^-(t+\tau)
\hat{\sigma}^-(t)\rangle }{[\langle \hat{\sigma}^+\hat{\sigma}^-\rangle (t)]^2 }
\end{equation}
where
\begin{equation}
\hat{\sigma}^-=\sum_l e^{-i\vec{k}\cdot\vec{R}_l}\hat{\sigma}_l^-=\hat{\sigma}^{+\dagger}
\end{equation}
which is the normalized intensity-intensity correlations of light
emitted in the $\hat{k}$ direction. The
denominator in Eq.~(\ref{Eqg2}) can be approximately calculated within
the operator method described above. However, calculating the
numerator requires an extension of the method.

The two time correlations can be computed from the expectation value of
operators using an algorithm based on a method for calculating the two
time expectation values from density matrices. For this discussion,
we will use the example in Eq.~(\ref{Eqg2}) for the two-time expectation
value. There are three steps in the calculation.
Step 1: 
First calculate the density matrix using Eq.~(\ref{EqDenMat}) up to time
$t$. Step 2: At this time, a projected density matrix is calculated using
\begin{equation}\label{EqDenMatNew}
\frac{\hat{\sigma}^-\hat{\rho}(t)\hat{\sigma}^+}{\langle \hat{\sigma}^+\hat{\sigma}^-\rangle (t)}\to\hat{\rho}(t)
\end{equation}
Step 3: The density matrix equation, Eq.~(\ref{EqDenMat}) is used to
propagate to time $t+\tau$. Finally, the two time expectation value is
given by
\begin{equation}
\frac{\langle \hat{\sigma}^+(t)\hat{\sigma}^+(t+\tau)\hat{\sigma}^-(t+\tau)
\hat{\sigma}^-(t)\rangle }{[\langle \hat{\sigma}^+\hat{\sigma}^-\rangle (t)]^2}= \frac{Tr[\hat{\sigma}^+\hat{\sigma}^-\hat{\rho}(t+\tau ))]}{\langle \hat{\sigma}^+\hat{\sigma}^-\rangle (t)}
\end{equation}

The operator method can be used within this scenario by tracking the
effect of the different steps. Since some of the equations are long
and complicated, Appendix~\ref{SecTtIC} contains the derivations
and the specific equations for both MF2 and MF3
calculations.

Step 1 is the same as usual where the
various coupled operator equations are solved as a function of time. The
expectation value is given by Eq.~(\ref{EqTrtA}) and is
\begin{equation}\label{EqTrt}
\langle \hat{\sigma}^+\hat{\sigma}^-\rangle (t)=\sum_n\left[\langle\hat{e}_n\rangle (t)
+\sum_m{\vphantom{\sum}}' e^{i\varphi_{mn}}
\langle\hat{\sigma}^+_m\hat{\sigma}^-_n\rangle (t)
\right]
\end{equation}
where $\varphi_{mn}=\vec{k}\cdot (\vec{R}_{m}-\vec{R}_n)$ and the
prime on the sum means $m\neq n$.

Step 2 does not change the trace of the density matrix {\it but} the 
expectation values change. The 
expectation values after the transformation, Eq.~(\ref{EqDenMatNew}),
can be found from the new density matrix. The part arising from
the numerator, $\hat{\sigma}^-\hat{\rho}(t)\hat{\sigma}^+$, are in
Appendix~\ref{SecTtIC}.

Step 3 is the same as Step 1 but using the initial conditions from
Step 2. At the time $t+\tau$, the expectation value
$\langle \hat{\sigma}^+\hat{\sigma}^-\rangle$ is evaluated
and is divided by the old value, $\langle \hat{\sigma}^+\hat{\sigma}^-\rangle (t)$,
to obtain $g^{(2)}(\tau )$.

\subsection{Restricting correlations}

The two atom contributions to the operator equations leads to an increasing
number of terms in the operator correlations: the time derivative of
the one atom expectation values depend on two atom expectation values,
the time derivative of
two atom expectation values depend on three atom expectation values,
etc. The set of equations only closes when the number of operators in
the expectation values equals the number of atoms.

The set of equations can be closed by using approximations for
the higher order expectation values. As described in Appendix~\ref{SecAppRed},
we obtain the equations by setting the cumulant\cite{RK11962}
at a given order to zero as done in
Ref.~\cite{FY11999,CYL2004,LY12012,KR12015,QR12019,OMG2019,HPR2020,KK12018,KRK2019,SSF2020}.
For example, the mean field (MF1) approximation
results by using the approximation of Eq.~(\ref{EqOp2}) to replace a
two atom expectation value by the product of one atom expectation
values as
\begin{equation}\label{EqMF1}
\langle \hat{Q}^j_n\hat{Q}^{j'}_{m}\rangle \gets 
\langle \hat{Q}^j_n\rangle \langle\hat{Q}^{j'}_{m}\rangle
\end{equation}

Typically, a better approximation results by keeping the two atom
expectation values but replacing the three atom expectation values by
their cumulant approximation, Eq.~(\ref{EqOp3}), to give
\begin{eqnarray}\label{EqMF2}
\langle \hat{Q}^j_n\hat{Q}^{j'}_{m}\hat{Q}^{j''}_{l}\rangle &\gets &
\langle \hat{Q}^j_n\hat{Q}^{j'}_{m}\rangle\langle\hat{Q}^{j''}_{l}\rangle +
\langle \hat{Q}^j_n\hat{Q}^{j''}_{l}\rangle\langle \hat{Q}^{j'}_{m}\rangle
\nonumber \\ &+&
\langle \hat{Q}^j_n\rangle\langle\hat{Q}^{j'}_{m}\hat{Q}^{j''}_{l}\rangle 
-2\langle \hat{Q}^j_n\rangle\langle\hat{Q}^{j'}_{m}\rangle\langle\hat{Q}^{j''}_{l}\rangle
\end{eqnarray}
Below, MF2 calculations are those where this replacement has been made.

An even better approximation typically results by keeping the
two and three operator expectation
values but approximating the four operator expectation values using
the replacement in Eq.~(\ref{EqOp4}). Below, MF3 calculations
are those where this replacement has been made.

This process can be continued to even higher order products which are typically
more accurate than those of lower order. In principle, going to higher order
is a systematic way for increasing the accuracy of a calculation. However, the
speed of the resulting programs rapidly decrease with the order kept.

\section{Results}\label{SecR}

In this section, we will refer to calculations that use Eq.~(\ref{EqMF1})
as mean field or MF1 calculations, those that
use Eq.~(\ref{EqMF2}) are MF2 calculations, and those that
use Eq.~(\ref{EqOp4}) are MF3 calculations.

\subsection{One-time expectation values}

This section describes results from three types of calculations that compare
the different order mean-field calculations
with that from a full density matrix treatment. The topics are presented
in order from least correlated to most correlated.

\subsubsection{Collective shift in a line of atoms}

We use the operator formalism to calculate an observable
from Ref.~\cite{GFS2020}. In Ref.~\cite{GFS2020}, they measured the collective
shift in scattered light from a one-dimensional chain of atoms. They
also performed calculations using the MF1 approximation, equivalent
to using Eq.~(\ref{EqMF1}), to interpret their results. For the most part,
there was good agreement between this approximation and the measurements
except for their Fig.~4(a) for the global resonance shift versus the
driving amplitude, $\Omega/\Gamma$. In this situation, they showed the
detuning versus the Rabi frequency and experimentally found a steep drop
in the shift at $\Omega\sim 1.5\Gamma$ whereas the calculation showed a
smooth decrease with increasing Rabi
frequency. They presented the comparison between
a full density matrix calculation and the MF1 approximation for
6 atoms which showed good agreement; this agreement suggests that the mean
field approximation itself was not the source of the disagreement. However,
it isn't completely clear whether or not the MF1 approximation could fail
at larger $N$ because the atoms are roughly in a line. We performed
calculations with many more atoms at both the MF1 and MF2
level as a test of the accuracy of the mean field approximation.

We constructed the simulation to mimic as many aspects of the experiment as
possible. The atoms are trapped using a Gaussian standing wave with, on
average, a 50\% probability for an atom at each intensity maximum. The atoms
are mainly along the $z$-axis and are excited with an $M=1$ transition.
The atoms have a Gaussian distribution in $x,y$ with a transverse width
$\sigma_\rho =300$~nm. The Gaussian beam has wavelength of 940~nm
and waist of 3.3~$\mu$m giving a Rayleigh range $Z_R=36.4$~$\mu$m.
The $z$-position of the atoms are found from the
zeros of $\sin [kz-\arctan (z/Z_R)]$ 
with $k=2\pi/(940\; {\rm nm})$. Since
the detector is perpendicular to the line of atoms, the measured light
scattered is proportional to $\sum_n\langle\hat{e}_n\rangle$.

\begin{figure}
\resizebox{80mm}{!}{\includegraphics{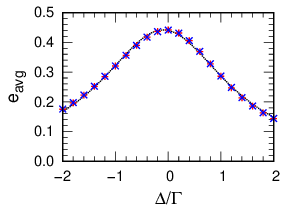}}
\caption{\label{FigGlic}
The average excitation probability for 100 atoms in the configuration of
Ref.~\cite{GFS2020} for $\Omega = 2\Gamma$.
The red $*$ are from a MF1 calculation while
the blue X are from a MF2 calculation. The black dotted line is
a Lorentzian fit to the MF2 points.
}
\end{figure}

Figure~\ref{FigGlic} shows results from a calculation with 100 atoms averaged
over 200 configurations of atom positions for $\Omega =2\Gamma$ where
the experiment has $\sim 0$ detuning. The red $*$ are from MF1 calculations while the blue X are from a MF2 calculation.
This is an ideal application of the higher order mean field calculations result because
we could increase the order until the results agree.
The black dotted line is a Lorentzian fit to the MF2
points. The fit gave a collective shift of $-0.093\Gamma$ which is in
line with the calculation in Ref.~\cite{GFS2020} but in disagreement with
the measurements. Thus, the MF1 approximation is not the cause for
disagreement in this experiment.


\begin{figure}
\resizebox{80mm}{!}{\includegraphics{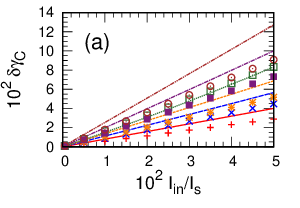}}
\resizebox{80mm}{!}{\includegraphics{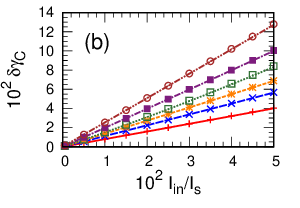}}
\resizebox{80mm}{!}{\includegraphics{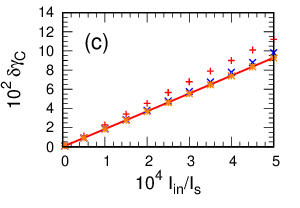}}
\caption{\label{FigMF}
The $\delta\gamma_C$ for the case of 7 equally spaced atoms, $\Delta z=0.4\lambda$,
excited to the different eigenmodes of light traveling along $z$ with
circular polarization. The $I_{in}/I_s=\Omega^2/(2\Gamma^2 )$.
In (a) and (b), the decay rate of the
eigenstate, $2{\rm Re}[G_\alpha ]$, corresponding to each
line in increasing order (solid red to dot-dash-dash brown) is 1.413, 1.294, 1.147, 0.975, 1.032, 0.961 times the one atom decay rate $\Gamma$. The
decay rate for (c) is 0.179 times $\Gamma$. In all cases, the lines
are from solving the full density matrix equations. The symbols in (a) are from
calculation using the approximation in Eq.~(\ref{EqMF1}) while those in
(b) are from Eq.~(\ref{EqMF2}). In both cases the color matches that of the
corresponding line (in order: red $+$, blue X, orange $*$, green hollow square,
purple filled square, brown hollow circle). In (c), the red $+$ is from
MF1,
Eq.~(\ref{EqMF1}), the blue X are from MF2,
Eq.~(\ref{EqMF2}), and the orange
$*$ are from MF3, Eq.~(\ref{EqOp4}).
}
\end{figure}

\subsubsection{Normal mode excitation}

As another example, we explore the case presented in Ref.~\cite{WR12020}
where a line of 7 atoms are excited with circularly polarized light.
We present results for the case where the atoms are separated by $0.4\lambda$.
Following Ref.~\cite{WR12020},
the atoms are excited with amplitudes taken from the eigenvectors,
$u_\alpha (\vec{R}_n)$,  of the
complex symmetric, $g^+_{mn}$: $\sum_n g^+_{mn}u_\alpha (\vec{R}_n)
=G_\alpha u_\alpha (\vec{R}_m)$ with the normalization
$\sum_n |u_\alpha (\vec{R}_n)|^2 = 1$\footnote{This follows the notation
in Ref.~\cite{WR12020}. The normalization
$\sum_n u_\alpha (\vec{R}_n)u_{\alpha '} (\vec{R}_n) = 1$ for complex
symmetric matrices
could have been chosen.}. To organize the states, we use the decay
rate of the eigenstate, $2{\rm Re}[G_\alpha ]$.
The $\Omega^+_n=\Omega u_\alpha (\vec{R}_m)$.
As with Ref.~\cite{WR12020}, we compute the rate of scattered photons
as
\begin{eqnarray}
\gamma&=&\sum_n\left[\Gamma\langle\hat{e}_n\rangle +\sum_m{\vphantom{\sum}}'\Gamma_{mn}\langle\hat{\sigma}^+_m\hat{\sigma}^-_n\rangle \right]
\label{EqDec}\\
\gamma_C &=&\sum_n\left[ \Gamma\langle\hat{\sigma}^+_n\rangle\langle\hat{\sigma}^-_n\rangle +\sum_m{\vphantom{\sum}}'\Gamma_{mn}\langle\hat{\sigma}^+_m\rangle\langle
\hat{\sigma}^-_n\rangle  \right] \\
\gamma_I &=& \gamma-\gamma_C
\end{eqnarray}
where $\Gamma_{mn}$ is from Eq.~(\ref{Eqgdef1}) and a prime on the sum
indicates ${m\neq n}$.
The $\gamma$ is the total scattering
rate; the $\gamma_C$ is the classical approximation to the scattering rate; 
the $\gamma_I$ is the incoherent scattering rate.

As with Ref.~\cite{WR12020}, we calculate $\delta \gamma_C\equiv (\gamma_C^{\rm lin}
-\gamma_C)/\gamma_C$ and the $\gamma_I/\gamma_C$ for the different eigenstates.
The $\gamma_C^{\rm lin}$ uses the linear approximation to the operator
equations; this is equivalent to using Eq.~(\ref{EqMF1}) with the
additional approximation $\langle \hat{e}_n\rangle(t)=0$.
For the intensities investigated, the $\gamma_C^{\rm lin}\sim\gamma_C$ which
means the $\delta \gamma_C$ strongly emphasizes any error in the approximation.

Figure~\ref{FigMF} shows plots versus the incident intensity in
units of the saturation intensity, $I_{in}/I_s\equiv 2\Omega^2/\Gamma^2 $,
comparing the density matrix results (lines in all figures)
to those using the MF1 approximation
(Fig.~\ref{FigMF}(a) and red $+$ in (c))
and MF2 approximation
(Fig.~\ref{FigMF}(b) and blue X in (c)). 
The MF3 approximation is only plotted in 
Fig.~\ref{FigMF}(c) (orange $*$) because the MF2 is already
in agreement for the other cases.

As noted
in Ref.~\cite{WR12020}, the MF1 approximation, Eq.~(\ref{EqMF1}),
in Fig.~\ref{FigMF}(a) and red $+$ in (c)
has a qualitative relationship to the density matrix result.
This level of approximation does not even give the correct order to
the curves: the purple solid and green hollow squares are reversed.
However, the MF2 approximation, Eq.~(\ref{EqMF2}), provides
quantitative agreement with the density matrix result except for the most
subradiant state in Fig.~\ref{FigMF}(c). This is in agreement with the
finding in Ref.~\cite{KR12015} that the second order correlations greatly
improve the agreement with the full density matrix result. Lastly,
the MF3 approximation, Eq.~(\ref{EqOp4}), provides quantitative agreement
for all of the curves; this is shown in  Fig.~\ref{FigMF}(c) where this
approximation, the orange $*$, is on top of the red line for all intensities.
To give an idea of the accuracy, the value for the largest intensity of
the most subradiant state has
an error of 18\%, 5.0\%, and 0.18\% for the MF1, MF2, and MF3
approximations respectively.

This is nearly an ideal application of the higher order mean field calculations result because
we could increase the order until the results agree for all but one of the cases.
While the convergence with increasing order for the most subradiant case would
lead to the expectation that the MF3 result was accurate, it would be difficult
to quantify that accuracy without knowing the result from the exact, density matrix
calculation.

\subsubsection{Collective decay from the totally excited state}

The last example in this section is the case where there is no laser
causing stimulated absorption or emission
and all atoms start in the excited state.\cite{RHD1954}
This is the case treated by Dicke with the approximation of $\Omega_{mn}=0,
\Gamma_{mn}=\Gamma$
and can lead to superradiance or subradiance depending on the
circumstances. For our calculations, we will use the correct forms of
$\Omega_{mn}$ and $\Gamma_{mn}$.
Note that the MF1 equations lead to the
solution $\langle\hat{e}_n\rangle (t)=\exp (-\Gamma t)$ and
$\langle\hat{\sigma }^\pm_n\rangle (t)=0$ which does contain any
multiatom effects.

\begin{figure}
\resizebox{80mm}{!}{\includegraphics{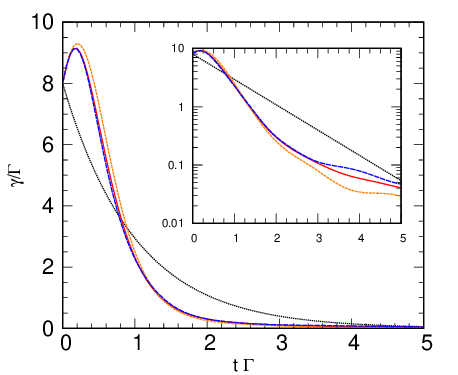}}
\caption{\label{FigDicke}
The instantaneous decay rate $\gamma$, Eq.~(\ref{EqDec}), for 8
equally spaced atoms with a separation: $\Delta z =0.1\lambda$.
The states only support $M=1$ transitions. The red solid line is from
solving the density matrix equations. The orange short-dashed line is from
solving the operator equations at the MF2, Eq.~(\ref{EqMF2}),
level and the dashed blue line is at the MF3, Eq.~(\ref{EqOp4}),
level. The black dotted line is $8e^{-\Gamma t}$ and is the MF1 solution.
The inset is the same plot but with a logarithmic scale for the
$y$-axis.
}
\end{figure}

For this calculation, we reproduce the case discussed in Ref.~\cite{MFO2020}
where atoms are equally spaced on a line.
Figure~\ref{FigDicke} shows the photon emission rate for 8 atoms
in a line with separation $\Delta z=0.1\lambda$ where the
states make $M=1$ transitions. Plotted are the emission rate for the density matrix
equations (red solid line), for the MF2 equations (blue dashed
line), for the MF3 approximation (orange short-dash line)
and the MF1 results
(black dotted line) which has no multi-atom effects. The early time increase
in photon emission rate is due to superradiance of this system which is
a collective effect. The MF1 calculation
can not reproduce any aspect of the collective decay. The
MF2 approximation reproduces many aspects of the time dependent
decay rate when the decay rate is large. In contrast, the MF3
approximation is in quantitative agreement with the exact result when the
decay rate is at least $\sim$1\% of the peak decay rate. At longer times
and smaller decay rates, none of the approximations are quantitatively
accurate as seen in the inset. This is during part of the time evolution
where subradiant states are the largest component of the signal,
indicating that the mean field approximations
are not accurately populating these states.

This is the worst case for one-time expectation values. The MF1 calculation
is not even qualitatively accurate, missing all of the important physics. The
MF2 and MF3 approximations decently agree with each other where $\gamma$ is
large. For this region, one would expect that the MF3 calculation is
accurate. However, for later times, the mean MF2 and MF3 calculations differ by
a factor of $\sim 2$ and, thus, one would not have any expectation for the level
of accuracy of the MF3. In fact, there are regions of substantial disagreement
with the exact, density matrix results indicating the subradiant states are not
accurately populated by the MF3 equations at late times.

\subsection{Two-time expectation values}

This section contains results from two-time expectation values. In particular,
we will examine how well the mean field approximations reproduce
the $g^{(2)}(\tau )$ function in Eq.~(\ref{Eqg2}) when the atoms are
uniformly excited as would happen in an experiment where atoms are
excited by a plane wave. We find that this is a more difficult test than
calculating one-time expectation values. For the cases we tested, the MF1 result
never gave an interesting result so we will only show the calculations using
the MF2 and MF3 approximations.

All of the calculations in this
section are for 7 atoms in a line separated by $0.4\lambda$ on the
$y$-axis. The incident light
is propagating perpendicular to the line of atoms in the $x$-direction.
Thus, the
$\Omega_n$ are identical. The atoms are linearly
polarized in the $z$-direction. We present results for
$g^{(2)}(\tau )$ for different $\hat{k}$ in the $xy$-plane. The
angle $\theta$ is the final direction of the photons relative to the
$x$-axis. The direction
$\theta =0$ has the most scattered photons and there is a rapid drop in
scattering probability as $\theta$ increases. In all of the
plots, the solid red line
is from a density matrix calculation, the short-dash orange line is from a 
MF2 calculation, and the dash blue line is from a MF3 calculation.

\begin{figure}
\resizebox{80mm}{!}{\includegraphics{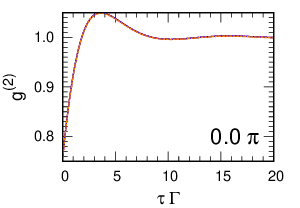}
\includegraphics{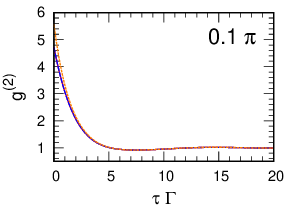}}
\resizebox{80mm}{!}{\includegraphics{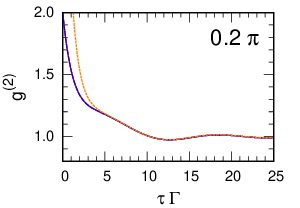}
\includegraphics{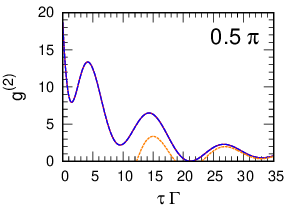}}
\caption{\label{Figg2_0.01}
The $g^{(2)}(\tau )$ from Eq.~(\ref{Eqg2}) for 7 atoms spaced by $0.4\lambda$ on the
$y$-axis with plane wave light incident from the $x$-direction. The atoms are linearly
polarized in the $z$-direction and the $\Omega_n = \Gamma /100$. The solid red line
is from a density matrix calculation, the short-dash orange line is from a 
MF2 calculation, and the dash blue line is from a mean MF3 calculation.
The calculations are for final photon angles of 0.0, 0.1, 0.2, and 0.5~$\pi$
relative to the $x$-axis in the $xy$-plane.
}
\end{figure}

Figure~\ref{Figg2_0.01} shows results for $g^{(2)}(\tau )$ for $\Omega_n =\Gamma /100$.
The one time average $\langle \hat{\sigma}^+\hat{\sigma}^-\rangle (t)$
from Eq.~(\ref{EqTrt}) at large time is proportional to the rate of photons
scattered into that direction. We did calculations of this expectation
value at angles from $0.0$ to $0.5\pi$ in
steps of $0.1\pi$ and found values of (3.11E-3, 1.08E-3, 1.38E-4,
5.88E-5, 5.84E-5, 3.20E-5) which shows the drop in photon scattering with
angle. Both the MF2 and MF3 calculations of these values
were accurate, with the mean MF2 error less than 0.1\% at $0.0\pi$ increasing
to 2\% error at $0.5\pi$
while the MF3 approximation was
accurate to at least 7 significant digits for all angles.
The $g^{(2)}(\tau )$ has antibunching behavior for scattering into $0.0\pi$
with $g^{(2)}(0)<1$
but has bunching behavior, $g^{(2)}(0)>1$,  for the larger angles calculated.
For photon scattering along the atom line, $0.5\pi$, the  $g^{(2)}(0)\simeq 19$.

The MF3 calculations reproduce the density matrix calculations for all
angles, even for larger angles where few photons are scattered
and where there is a strong correlation
between successive photons. This is not surprising; because the $\Omega_n$ are small,
the expectation values decrease rapidly with increasing number
of operators in the expectation value.
The MF2 calculation reproduces the density
matrix calculation for $0.0\pi$ and is in decent agreement for $0.1\pi$. However,
at $0.2\pi$ the $g^{(2)}(0)$ is too large by a factor of $\simeq 2.3$ and for larger
angles the error is even larger. Worse, the MF2 calculation is negative
for 0.4 and $0.5\pi$ which is unphysical. This is, perhaps, surprising because
$\Omega$ is small and an argument similar to that for MF3 can be made.
 Perhaps more interestingly, the MF2 approximation does
not increase in accuracy for the larger angles as $\Omega$ is decreased 
which violates the expectation that there should be less correlation
for smaller excitation probabilities: calculations with $\Omega =10^{-5}\Gamma$
had similar size errors to those in Fig.~\ref{Figg2_0.01}.
The difficulty is to get the correct value for $g^{(2)}(0)$. This incorrect initial
value means that the starting
conditions for the one- and two-atom operators are not accurate. The problem
appears to be with the $\langle\hat{\sigma}^+_m\hat{\sigma}^-_l\rangle (t=0)$
terms which are of the same size as the $\langle\hat{e}_l\rangle (t=0)$.
The starting two operator terms are evaluated from Eq.~(\ref{Eq2optt}) which
has three and four operator expectation values that are important. These expectation
values are
not included in the MF2 equations and must be approximated from the cumulant
relation.

\begin{figure}
\resizebox{80mm}{!}{\includegraphics{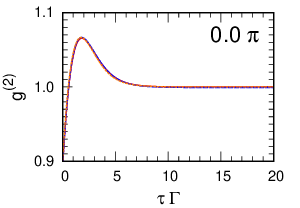}
\includegraphics{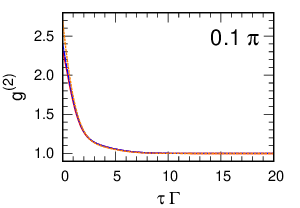}}
\resizebox{80mm}{!}{\includegraphics{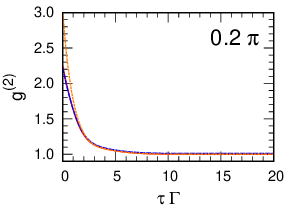}
\includegraphics{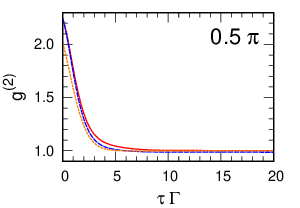}}
\caption{\label{Figg2_0.5}
Same as Fig.~\ref{Figg2_0.01} but for $\Omega_n = \Gamma /2$.
}
\end{figure}

Figure~\ref{Figg2_0.5} shows the same results for $\Omega_n =\Gamma /2$. For this
case, the one time average $\langle \hat{\sigma}^+\hat{\sigma}^-\rangle (t)$
from Eq.~(\ref{EqTrt}) at large time is (4.55, 0.480, 0.505, 0.389, 0.431, 0.452)
for the angles from $0.0$ to $0.5\pi$. The MF2 had 1\% error at
$0.0\pi$ increasing to 6\% error at $0.5\pi$ while the MF3 had less than
0.01\% error at $0.0\pi$ and 1\% error at $0.5\pi$. The $g^{(2)}(\tau )$
again shows antibunching behavior for $0.0\pi$ but bunching for larger angles.
Unlike the previous weak laser case, the $g^{(2)}(\tau )$ does not achieve
very large values for the larger angles. This is due to the higher intensity
leading to more excitation and less suppression of photon scattering into the
higher angles. Also unlike the previous case, the MF2 does not have
qualitative errors which indicates there
is less correlation. The MF3 does have
visible errors for the larger angles indicating the start of the breakdown
of this approximation, but, the errors are small.

A more interesting aspect of the breakdown was that
the $g^{(2)}(\tau )$ did not go to 1 at large $\tau$ for MF3. This indicates
the errors are large enough to trap the non-linear MF3 equations in
unphysical parameter regions for the expectation values. We examined the
$\langle \hat{e}_l\hat{e}_m\rangle$ after doing the projection step, Eq.~(\ref{EqDenMatNew}).
For the weak laser case, $\Omega_n=\Gamma /100$. Both mean field approximations
had instances of negative values which is not physical but the MF3
was much smaller. For MF3, the negative values were $\sim 100\times$
smaller in magnitude than the largest positive values while the negative values were
only $\sim 5\times$ smaller for MF2. There were no negative values
for the $\Omega_n=\Gamma /2$ example because there is relatively more photon scattering
into the larger angles.
This breakdown is a reminder that physical
constraints are not necessarily obeyed in the mean field approximations.

This case shows that it will often be difficult to show convergence of the mean field
equations with increasing order for the two-time expectation values. Although
the MF3 was in good agreement with the density matrix results for all
cases, some external test of convergence may be necessary. This external test might
be comparison with experiment or comparison with density matrix results for smaller
number of atoms.

\subsection{Unphysical behavior}

The previous section had examples where the MF2 and MF3 results were not
just incorrect but were unphysical for some parameters. In other
situations, it was found that the higher order mean field equations were
unstable\cite{OSW2017,CGO2018} which was not the case for any of the
examples we investigated. It could be that the damping Eq.~\eqref{EqDenMat}
helps stabilize the higher order mean field. Another possibility is
the atomic arrays prevent large values of $g^\pm_{nm}$ that occur with
randomly placed atoms; perhaps large $g^\pm_{nm}$ could lead
to instability. If instabilities occur, they would be due to
the inherent inaccuracy of the equations and could not be fixed by,
for example, using smaller $\delta t$ when solving the coupled equations.

\section{Summary}\label{SecS}

Higher order mean field calculations of both one- and two-time expectation values
were calculated for the interaction of
light with 2-level atoms in four different situations. For the one-time
expectation values, we have repeated the derivation of Ref.~\cite{KR12015} but
for the complex expectation values from $\hat{\sigma}^\pm$ instead of the
real expectation values from the Pauli spinors.
As higher order expectation values are
included, the mean field method becomes more accurate but more difficult
to implement. We have also derived the equations
needed to compute the two-time expectation values, for example $g^{(2)}(\tau )$.
We found the two-time expectation values to be less accurate than one-time
expectation values for a given
mean field order. The calculations also highlight the difficulty,
in some cases, of proving convergence by increasing the order of the mean field
equations.

The one-time expectation values were most easily reproduced using the higher order
mean field approximations.
The first case addressed was the experiment of Ref.~\cite{GFS2020} which measured
the shift in the scattering of light from a line of atoms. The
MF1 and MF2 results were nearly identical. The next case was based
on Ref.~\cite{WR12020} which examined the scattering of light from a line of
atoms when the excitation pattern was an eigenmode of the interaction.
While the MF1 results were in qualitative agreement with exact
calculations, the
MF2 and MF3 results were in better agreement so that the error was
less then 0.2\% for all cases at the MF3 level. 
The last case was the collective decay from a line of atoms that start with
all atoms excited. The MF1 result was not even in qualitative agreement;
it could not account for even the basic physics. The MF2 was in
decent agreement when the photon emission rate was large and the MF3
was in excellent agreement in this range. However, at later times, the
MF2 was more than a factor of 2 off while the MF3 was in rough
agreement indicating neither approximation was accurate with regards to the population
of subradiant states at late times.

The two-time expectation values were represented by calculations of $g^{(2)}
(\tau )$ for
a line of atoms excited by a plane wave. The MF1 results were
qualitatively incorrect.
For the angles where the majority of photons were scattered, the MF2 and
MF3 approximations reproduced the time dependence of $g^{(2)}
(\tau )$. However, at angles
where few photons were scattered, the MF2 calculation was not even in
qualitative agreement and in some cases had strongly unphysical features. The MF3
approximation was in excellent agreement for weak light and in good agreement for
stronger excitation in the cases tested. In some cases, an external method for
estimating the accuracy of the mean field method will be needed.

Overall, it appears that the MF2 approximation will often be accurate for
calculations of one-time expectation values. However, for many two-time expectation
values, the MF3 approximation will be needed.

Just before submitting this paper, we became aware of Ref.~\cite{PHR2021}
which describes a computational toolbox for the cases discussed in our
paper as well as extensions for cavities and mechanical oscillation
of a membrane.

\begin{acknowledgments}
We thank J. Ruostekoski and L.A. Williamson for helping
us understand Ref.~\cite{WR12020}. This work was supported by the National Science Foundation under Grant
No. 1804026-PHY.
\end{acknowledgments}

\appendix

\section{Reduction of operator expectation values}\label{SecAppRed}

This section contains the algorithm used to reduce higher order
expectation values to lower order.

\subsection{Cumulants for 2 through 5 operators}\label{SecAppCum}

The operator equations do not close until the number of operators equals
the number of atoms. To obtain useful equations when the number of operators
is much less than the number of atoms, a method for approximating a
higher order product of operators by those with smaller products must
be chosen. The results presented here use a cumulant\cite{RK11962} based method for
the approximation. The Nth order cumulant is defined as
\begin{equation}
\kappa (\hat{Q}_1,...,\hat{Q}_N)=\sum_\pi  (-1)^{|\pi|-1}(|\pi|-1)!
\prod_{B\in \pi}\langle \prod_{i\in B}\hat{Q}_i\rangle
\end{equation}
where $\pi$ is the list of all partitions, $B$ is all blocks of partition $\pi$,
and $|\pi|$ is the number of parts. For example, for 4 operators, there are
4 permutations with $|\pi |=2$ parts with the expectation value of 3 operators
times the expectation value of 1 operator:
$\langle \hat{Q}_a\hat{Q}_b \hat{Q}_c\rangle \langle \hat{Q}_d\rangle$,
$\langle \hat{Q}_a\hat{Q}_b \hat{Q}_d\rangle \langle \hat{Q}_c\rangle$,
$\langle \hat{Q}_a\hat{Q}_c \hat{Q}_d\rangle \langle \hat{Q}_b\rangle$,
$\langle \hat{Q}_b\hat{Q}_c \hat{Q}_d\rangle \langle \hat{Q}_a\rangle$.
As another example,
for 4 operators, there are two forms of the case with $|\pi | =2$: the
expectation value of 3 operators times the expectation value of 1 operator
with 4 permutations and the expectation value of 2 operators times the expectation
value of 2 operators with 3 permutations.

All operators in the expectation values commute in the equations above
because they act on different atoms. This leads to the cumulants
\begin{equation}\label{EqCu2}
\kappa (\hat{A},\hat{B})
=\langle\hat{A}\hat{B}\rangle-\langle\hat{A}\rangle
\langle\hat{B}\rangle
\end{equation}
\begin{eqnarray}\label{EqCu3}
\kappa (\hat{A},\hat{B},\hat{C})
&=&\langle \hat{A}\hat{B}\hat{C}\rangle -
(\langle \hat{A}\hat{B}\rangle\langle \hat{C}\rangle + \langle\hat{A}\hat{C}\rangle\langle \hat{B}\rangle\nonumber\\
&\null&  +
\langle\hat{B}\hat{C}\rangle\langle \hat{A}\rangle)
+2\langle \hat{A}\rangle\langle\hat{B}\rangle\langle \hat{C}\rangle
\end{eqnarray}
\begin{eqnarray}\label{EqCu4}
\kappa 
&=&\langle  \hat{A}\hat{B}\hat{C}\hat{D}\rangle -
(\langle  \hat{A}\hat{B}\hat{C}\rangle\langle\hat{D}\rangle
+3\Pi )\nonumber\\
&-& (\langle \hat{A}\hat{B}\rangle\langle\hat{C}\hat{D}\rangle
+2\Pi) +2(\langle \hat{A}\hat{B}\rangle\langle\hat{C}\rangle\langle\hat{D}\rangle
+5\Pi)
\nonumber\\
& -&6\langle \hat{A}\rangle\langle\hat{B}\rangle\langle\hat{C}\rangle\langle\hat{D}\rangle
\end{eqnarray}
\begin{eqnarray}\label{EqCu5}
\kappa &=&\langle  \hat{A}\hat{B}\hat{C}\hat{D}\hat{E}\rangle -
(\langle  \hat{A}\hat{B}\hat{C}\hat{D}\rangle\langle\hat{E}\rangle
+4\Pi)\nonumber\\
&-&(\langle  \hat{A}\hat{B}\hat{C}\rangle\langle\hat{D}\hat{E}\rangle +9\Pi)
+2(\langle  \hat{A}\hat{B}\hat{C}\rangle\langle\hat{D}\rangle\langle\hat{E}\rangle
+9\Pi )\nonumber\\
&+&2(\langle  \hat{A}\hat{B}\rangle\langle\hat{C}\hat{D}\rangle\langle\hat{E}\rangle
+14\Pi)-6(\langle \hat{A}\hat{B}\rangle\langle\hat{C}\rangle\langle\hat{D}\rangle\langle\hat{E}\rangle
+9\Pi)\nonumber\\
&+&24\langle\hat{A}\rangle\langle\hat{B}\rangle\langle\hat{C}\rangle\langle\hat{D}\rangle\langle\hat{E}\rangle
\end{eqnarray}
where the $N\Pi$ in these equations indicate the $N$ permutations with the same
form as the first term in that parenthesis.

\subsection{Replacement of higher order expectation values}

This paper truncates the order of the equations by replacing higher order
operators by products of lower order operators. The prescription used
is to set the cumulant to 0. The mean field (MF1) equations
arise by setting $\kappa =0$ in Eq.~(\ref{EqCu2}) giving
\begin{equation}\label{EqOp2}
\langle\hat{A}\hat{B}\rangle \gets \langle\hat{A}\rangle
\langle\hat{B}\rangle
\end{equation}
The MF2 equations arise by setting $\kappa =0$ in Eq.~(\ref{EqCu3}) giving
\begin{equation}\label{EqOp3}
\langle \hat{A}\hat{B}\hat{C}\rangle \gets
\langle \hat{A}\hat{B}\rangle\langle \hat{C}\rangle + \langle\hat{A}\hat{C}\rangle\langle \hat{B}\rangle  +
\langle\hat{B}\hat{C}\rangle\langle \hat{A}\rangle
-2\langle \hat{A}\rangle\langle\hat{B}\rangle\langle \hat{C}\rangle
\end{equation}
The MF3 equations arise by setting $\kappa =0$ in Eq.~(\ref{EqCu4}) giving
\begin{eqnarray}\label{EqOp4}
&\null&\langle  \hat{A}\hat{B}\hat{C}\hat{D}\rangle \gets 
(\langle  \hat{A}\hat{B}\hat{C}\rangle\langle\hat{D}\rangle
+3\Pi )
+ (\langle \hat{A}\hat{B}\rangle\langle\hat{C}\hat{D}\rangle
+2\Pi)\nonumber\\ &\null&\qquad -2(\langle \hat{A}\hat{B}\rangle\langle\hat{C}\rangle\langle\hat{D}\rangle
+5\Pi)
 +6\langle \hat{A}\rangle\langle\hat{B}\rangle\langle\hat{C}\rangle\langle\hat{D}\rangle
\end{eqnarray}

In the two-time formalism, the initial conditions contain expectation values
for 2 more operators than are being propagated. For example, the MF2
initial conditions contain expectation values of 4 operators. Thus,
Eq.~(\ref{EqOp4})
can not be directly used because it contains expectation values of 3
operators. For this case, we replace the expectation value of 3 operators
with the expression in Eq.~(\ref{EqOp3}). This leads to the replacement
in the MF2 initial conditions
\begin{equation}\label{EqTt2}
\langle  \hat{A}\hat{B}\hat{C}\hat{D}\rangle \gets (
\langle  \hat{A}\hat{B}\rangle\langle\hat{C}\hat{D}\rangle +2\Pi)
-2\langle  \hat{A}\rangle\langle\hat{B}\rangle\langle\hat{C}\rangle\langle\hat{D}\rangle
\end{equation}
Similarly, there are expectation values of 5 operators in the two time
initial conditions at the MF3 level. We have not worked out
the resulting equation. Instead, in our simulation, we called the
expression in Eq.~(\ref{EqOp4}) when evaluating the 4 operator expectation
value in the expression
\begin{eqnarray}
&\null&\langle \hat{A}\hat{B}\hat{C}\hat{D}\hat{E}\rangle \gets
(\langle  \hat{A}\hat{B}\hat{C}\hat{D}\rangle\langle\hat{E}\rangle
+4\Pi)
+(\langle  \hat{A}\hat{B}\hat{C}\rangle\langle\hat{D}\hat{E}\rangle +9\Pi)\nonumber\\
&\null&\qquad -2(\langle  \hat{A}\hat{B}\hat{C}\rangle\langle\hat{D}\rangle\langle\hat{E}\rangle
+9\Pi )
-2(\langle  \hat{A}\hat{B}\rangle\langle\hat{C}\hat{D}\rangle\langle\hat{E}\rangle
+14\Pi)\nonumber\\
&\null&\quad+6( \hat{A}\hat{B}\rangle\langle\hat{C}\rangle\langle\hat{D}\rangle\langle\hat{E}\rangle
+9\Pi)
+24\langle\hat{A}\rangle\langle\hat{B}\rangle\langle\hat{C}\rangle\langle\hat{D}\rangle\langle\hat{E}\rangle
\end{eqnarray}

\section{Two time initial conditions}\label{SecTtIC}

For the two time expectation values, the density matrix is reset due to the
measurement at time $t$. An example case is given in Eq.~(\ref{EqDenMatNew}).
This leads to a reset
in the values of the operators. This section gives a derivation of the
new terms in the equation and the new operator values for the
example in Eq.~(\ref{EqDenMatNew}). The main complication is from the
numerator so this section only discusses the change in operators for
a new density matrix 
\begin{equation}\label{EqDenMatNewN}
\hat{\sigma}^-\hat{\rho}(t)\hat{\sigma}^+\to\hat{\rho}.    
\end{equation}

All of the quantities of interest can be cast as finding the expectation value
of any operator $\hat{A}$ after the transformation in Eq.~(\ref{EqDenMatNewN}).
This can be written as
\begin{eqnarray}\label{EqTtA}
\langle\hat{A}\rangle &=& Tr[\hat{A}\hat{\sigma}^-\hat{\rho}(t)\hat{\sigma}^+]
=\langle\hat{\sigma}^+ \hat{A}\hat{\sigma}^-\rangle \nonumber\\
&=&
\sum_{l,m} e^{i\varphi_{ml}}\langle\hat{\sigma}^+_m\hat{A}\hat{\sigma}^-_l\rangle
\end{eqnarray}
where $\varphi_{ml}=\vec{k}\cdot (\vec{R}_m-\vec{R}_l)$. The complications
arise when the operator $\hat{A}$ and one of the $\hat{\sigma}$ operators are for
the same atom. For this case, the relations
$\hat{e}_n\hat{\sigma}_n^-=0$, $\hat{\sigma}_n^+\hat{e}_n=0$, and
$\hat{\sigma}_n^+\hat{\sigma}_n^-=\hat{e}_n$ are useful.

The trace of the density matrix in Eq.~(\ref{EqDenMatNewN})
gives $\langle \hat{\sigma}^+\hat{\sigma}^-\rangle (t)$
and is obtained by setting $\hat{A}=\hat{1}$ in Eq.~(\ref{EqTtA}) giving
\begin{equation}\label{EqTrtA}
Tr[\hat{\rho}]_t=\sum_l\left[\langle\hat{e}_l\rangle (t)
+\sum_m{\vphantom{\sum}}' e^{i\varphi_{ml}}
\langle\hat{\sigma}^+_{m}\hat{\sigma}^-_l\rangle (t)
\right]
\end{equation}
where the prime on the sum indicates the summation does not include identical
indices, i.e. $m\neq l$.

The expressions for one atom expectation values after the transformation,
Eq.~(\ref{EqDenMatNewN}), involve two and three operator
expectation values:
\begin{eqnarray}
\langle\hat{\sigma}^-_n\rangle&=&\sum_l{\vphantom{\sum}}' [\langle\hat{\sigma}^-_n
\hat{e}_l\rangle +e^{i\varphi_{nl}}\langle\hat{e}_n\hat{\sigma}^-_l
\rangle \nonumber\\
&\null &+\sum_m{}{\vphantom{\sum}}' e^{i\varphi_{ml}}
\langle\hat{\sigma}^-_n\hat{\sigma}^+_m
\hat{\sigma}^-_{l}\rangle ] \\
\langle\hat{e}_n\rangle&=&\sum_l{\vphantom{\sum}}' [\langle\hat{e}_n
\hat{e}_l\rangle 
+\sum_m{\vphantom{\sum}}' e^{i\varphi_{ml}}
\langle\hat{e}_n\hat{\sigma}^+_{m}
\hat{\sigma}^-_{l}\rangle ]
\end{eqnarray}
where the prime on a sum means the summation does not include identical
indices (in each equation, the first sum has $l\neq n$ and the second sum
has $m\neq l$ and $m\neq n$). The
$\langle \hat{\sigma}^+_n\rangle=\langle\hat{\sigma}^-_n\rangle^*$.

In these equations and the ones below, the left hand side is the expectation
value for the operators after the transformation in Eq.~(\ref{EqDenMatNewN})
while the right hand side is before the transformation. This can be thought
of as obtaining the expectation value
at time $t$ {\it after} the
measurement whereas those on the right hand
side are at time $t$ {\it before} the measurement. For simplicity of notation,
the $(t)$ is dropped from this and the following equations.

The expressions for the two atom expectation values are only for $n\neq o$
and involve three and four operator
expressions:
\begin{eqnarray}
\langle\hat{\sigma}^-_n\hat{\sigma}^-_o\rangle &=&\sum_l{\vphantom{\sum}}' [
\langle\hat{\sigma}^-_n\hat{\sigma}^-_o\hat{e}_l\rangle
+e^{i\varphi_{nl}}\langle\hat{e}_n\hat{\sigma}^-_o\hat{\sigma}^-_l\rangle
\label{EqTtmm}
\nonumber\\
&+&e^{i\varphi_{ol}}\langle\hat{\sigma}^-_n\hat{e}_o\hat{\sigma}^-_l\rangle
+\sum_m{\vphantom{\sum}}'
e^{i\varphi_{ml}}\langle\hat{\sigma}^-_n\hat{\sigma}^-_o\hat{\sigma}^+_m
\hat{\sigma}^-_l\rangle
]\\
\langle\hat{\sigma}^-_n\hat{e}_o\rangle &=&\sum_l{\vphantom{\sum}}' [
\langle\hat{\sigma}^-_n\hat{e}_o\hat{e}_l\rangle
+e^{i\varphi_{nl}}\langle\hat{e}_n\hat{e}_o\hat{\sigma}^-_l\rangle
\nonumber\\
&\null&+\sum_m{\vphantom{\sum}}'
e^{i\varphi_{ml}}\langle\hat{\sigma}^-_n\hat{e}_o\hat{\sigma}^+_m
\hat{\sigma}^-_l\rangle
]\\
\langle\hat{e}_n\hat{e}_o\rangle &=&\sum_l{\vphantom{\sum}}' [
\langle\hat{e}_n\hat{e}_o\hat{e}_l\rangle +\sum_m{\vphantom{\sum}}'
e^{i\varphi_{ml}}\langle\hat{e}_n\hat{e}_o\hat{\sigma}^+_m
\hat{\sigma}^-_l\rangle
]\\
\langle\hat{\sigma}^-_n\hat{\sigma}^+_o\rangle &=&
e^{i\varphi_{no}}\langle\hat{e}_n\hat{e}_o\rangle +
\sum_l{\vphantom{\sum}}' [
\langle\hat{\sigma}^-_n\hat{\sigma}^+_o\hat{e}_l\rangle
+e^{i\varphi_{nl}}\langle\hat{e}_n\hat{\sigma}^+_o\hat{\sigma}^-_l\rangle
\nonumber\\
&+&e^{i\varphi_{lo}}\langle\hat{\sigma}^-_n\hat{e}_o\hat{\sigma}^+_l\rangle
+\sum_m{\vphantom{\sum}}'
e^{i\varphi_{ml}}\langle\hat{\sigma}^-_n\hat{\sigma}^+_o\hat{\sigma}^+_m
\hat{\sigma}^-_l\rangle\label{Eq2optt}
]
\end{eqnarray}
where the prime on a sum means the summation does not include identical
indices (in each equation, the first sum has $l\neq n,o$ and the second sum
has $m\neq n,o,l$).
All other expectation values can be obtained by complex conjugating these
(e.g. $\langle\hat{\sigma}^+_n\hat{e}_m\rangle
=\langle\hat{\sigma}^-_n\hat{e}_m\rangle^*$). While these equations look
complicated, the expressions come from straightforward application
of Eq.~(\ref{EqTtA}). For example, in Eq.~(\ref{EqTtmm}), the first
term arises when $m=l$ but $l\neq n$ or $o$, the second term is when
$m=n$ but $l\neq n$ or $o$, the third term is when
$m=o$ but $l\neq n$ or $o$, and the last term is when none of the
indices are the same.

The expressions for the three atom expectation values are only for the case where
none of $n,o,p$ are the same and involve four and five operator
expressions:
\begin{eqnarray}
\langle\hat{\sigma}^-_n\hat{\sigma}^-_o\hat{\sigma}^-_p\rangle
&=&\sum_l{\vphantom{\sum}}' [
\langle\hat{\sigma}^-_n\hat{\sigma}^-_o\hat{\sigma}^-_p\hat{e}_l\rangle
+e^{i\varphi_{nl}}\langle\hat{e}_n\hat{\sigma}^-_o\hat{\sigma}^-_p\hat{\sigma}^-_l\rangle
\nonumber\\
&+&e^{i\varphi_{ol}}\langle\hat{\sigma}^-_n\hat{e}_o\hat{\sigma}^-_p\hat{\sigma}^-_l\rangle
+e^{i\varphi_{pl}}\langle\hat{\sigma}^-_n\hat{\sigma}^-_o\hat{e}_p\hat{\sigma}^-_l\rangle\nonumber\\
&+&\sum_m{\vphantom{\sum}}'
e^{i\varphi_{ml}}\langle\hat{\sigma}^-_n\hat{\sigma}^-_o\hat{\sigma}^-_p\hat{\sigma}^+_m
\hat{\sigma}^-_l\rangle
]
\end{eqnarray}
\begin{eqnarray}
\langle\hat{\sigma}^-_n\hat{\sigma}^-_o\hat{e}_p\rangle
&=&\sum_l{\vphantom{\sum}}' [
\langle\hat{\sigma}^-_n\hat{\sigma}^-_o\hat{e}_p\hat{e}_l\rangle
+e^{i\varphi_{nl}}\langle\hat{e}_n\hat{\sigma}^-_o\hat{e}_p\hat{\sigma}^-_l\rangle
\nonumber\\
&+&e^{i\varphi_{ol}}\langle\hat{\sigma}^-_n\hat{e}_o\hat{e}_p\hat{\sigma}^-_l\rangle
\nonumber\\
&+&\sum_m{\vphantom{\sum}}'
e^{i\varphi_{ml}}\langle\hat{\sigma}^-_n\hat{\sigma}^-_o\hat{e}_p\hat{\sigma}^+_m
\hat{\sigma}^-_l\rangle
]
\end{eqnarray}
\begin{eqnarray}
\langle\hat{\sigma}^-_n\hat{e}_o\hat{e}_p\rangle
&=&\sum_l{\vphantom{\sum}}' [
\langle\hat{\sigma}^-_n\hat{e}_o\hat{e}_p\hat{e}_l\rangle
+e^{i\varphi_{nl}}\langle\hat{e}_n\hat{e}_o\hat{e}_p\hat{\sigma}^-_l\rangle
\nonumber\\
&+&\sum_m{\vphantom{\sum}}'
e^{i\varphi_{ml}}\langle\hat{\sigma}^-_n\hat{e}_o\hat{e}_p\hat{\sigma}^+_m
\hat{\sigma}^-_l\rangle
]
\end{eqnarray}
\begin{eqnarray}
\langle\hat{e}_n\hat{e}_o\hat{e}_p\rangle
&=&\sum_l{\vphantom{\sum}}' [
\langle\hat{e}_n\hat{e}_o\hat{e}_p\hat{e}_l\rangle
\nonumber\\
&+&\sum_m{\vphantom{\sum}}'
e^{i\varphi_{ml}}\langle\hat{e}_n\hat{e}_o\hat{e}_p\hat{\sigma}^+_m
\hat{\sigma}^-_l\rangle
]\\
\langle\hat{\sigma}^-_n\hat{\sigma}^-_o\hat{\sigma}^+_p\rangle
&=&e^{i\varphi_{np}}\langle\hat{e}_n\hat{\sigma}^-_o\hat{e}_p\rangle
+e^{i\varphi_{op}}\langle\hat{\sigma}^-_n\hat{e}_o\hat{e}_p\rangle \nonumber\\
&+&\sum_l{\vphantom{\sum}}' [
\langle\hat{\sigma}^-_n\hat{\sigma}^-_o\hat{\sigma}^+_p\hat{e}_l\rangle
+e^{i\varphi_{nl}}\langle\hat{e}_n\hat{\sigma}^-_o\hat{\sigma}^+_p\hat{\sigma}^-_l\rangle
\nonumber\\
&+&e^{i\varphi_{ol}}\langle\hat{\sigma}^-_n\hat{e}_o\hat{\sigma}^+_p\hat{\sigma}^-_l\rangle
+e^{-i\varphi_{pl}}\langle\hat{\sigma}^-_n\hat{\sigma}^-_o\hat{e}_p\hat{\sigma}^+_l\rangle\nonumber\\
&+&\sum_m{\vphantom{\sum}}'
e^{i\varphi_{ml}}\langle\hat{\sigma}^-_n\hat{\sigma}^-_o\hat{\sigma}^+_p\hat{\sigma}^+_m
\hat{\sigma}^-_l\rangle
]\\
\langle\hat{\sigma}^-_n\hat{e}_o\hat{\sigma}^+_p\rangle
&=&e^{i\varphi_{np}}\langle\hat{e}_n\hat{e}_o\hat{e}_p\rangle
+\sum_l{\vphantom{\sum}}' [
\langle\hat{\sigma}^-_n\hat{e}_o\hat{\sigma}^+_p\hat{e}_l\rangle\nonumber \\
&+&e^{i\varphi_{nl}}\langle\hat{e}_n\hat{e}_o\hat{\sigma}^+_p\hat{\sigma}^-_l\rangle
+e^{-i\varphi_{pl}}\langle\hat{\sigma}^-_n\hat{e}_o\hat{e}_p\hat{\sigma}^+_l\rangle\nonumber\\
&+&\sum_m{\vphantom{\sum}}'
e^{i\varphi_{ml}}\langle\hat{\sigma}^-_n\hat{e}_o\hat{\sigma}^+_p\hat{\sigma}^+_m
\hat{\sigma}^-_l\rangle
]
\end{eqnarray}
where the prime on a sum means the summation does not include identical
indices (in each equation, the first sum has $l\neq n,o,p$ and the second sum
has $m\neq n,o,p,l$).
All other expectation values can be obtained by complex conjugating these.

\bibliography{oper_meth}

\end{document}